\documentclass[twocolumn, useAMS]{mn2e}
\usepackage{times}
\usepackage{graphicx}
\usepackage{amsfonts}
\usepackage{subfigure}
\usepackage{color}

\usepackage[active]{srcltx}

\usepackage{ifpdf}
\ifpdf
\usepackage[hyperfootnotes=false,colorlinks=true]{hyperref}
\hypersetup{linkcolor=red,citecolor=blue}
\usepackage{epstopdf}
\fi

\newcommand{\al}{\alpha}

\newcommand{\f}{\frac}

\def\be {\begin{equation}}
\def\ee {\end{equation}}

\begin{document}
\title[Refined masses of 12 pulsars ...]{Strange star equation of state fits the refined mass measurement of 12 pulsars and predicts their radii }
\author[Gangopadhyay et al.]{Taparati Gangopadhyay $^{1}$, Subharthi Ray $^{2}$, Xiang-Dong Li $^{3}$, Jishnu Dey $^{1, \dagger}$, \\
\newauthor Mira Dey $^{1, \dagger}$\\
\\ $^1$ Dept. of Physics, Presidency
University, 86/1 , College Street, Kolkata 700 073, India.
\\$^2$ Astrophysics and Cosmology Research Unit, School of Mathematics, Statistics and Computer Sciences,  \\
Univ. of KwaZulu-Natal, Private Bag X54001, Durban 4000, South Africa
\\$^3$ Department of Astronomy, Nanjing University, and Key Laboratory of Modern Astronomy and Astrophysics, \\ 
Ministry of Education, Nanjing 210093, P. R. China
\\$^\dagger$ Associate of IUCAA, Pune. 
}

\maketitle

\begin{abstract}{There are three categories of stars whose masses have been found accurately in recent times: 
(1) two for which Shapiro delay is used which is possible due to GR light bending as the partner is heavy : PSR J1614 $-$ 2230
 and PSR J1903+0327 (2) six eclipsing stars for which numerical Roche Lobe geometry is used and (3) 3 stars for which 
spectroscopic methods are used and in fact for these three the mass and radii both are estimated. Motivated by large color ($N_c$ )
expansion using a modified Richardson potential, along with density dependent quark masses
 thereby allowing chiral symmetry restoration, we get compact strange stars fitting all the observed masses.}
\end{abstract} 

\vskip .5cm

\noindent Keywords(pacs) {stars:pulsars:general} {Equations of state} {stars:neutron}

\vskip 0.5cm

\section{Introduction}

High masses were found in different types of pulsar binaries, starting with some in globular clusters 
(Ransom \emph{et al.} 2005; Freire \emph{et al.} 2008a, 2008b). In these cases, however, the masses rely on observations
of periastron advance, which is assumed to be due to general relativistic effects only (rather than classical ones such as
due to rotationally and tidally induced quadrupoles), and statistical arguments that the inclinations are unlikely to be very
low. Thus, it was still possible to doubt that very massive neutron stars could exist. However, such doubts disappeared, with 
accurate mass determinations for PSR~J1614$+$2230 $(1.97 \pm 0.04M_\odot)$ (Demorest \emph{et al.} 2010) and
PSR J1903+0327 $(1.67 \pm 0.02M\odot)$ (Freire \emph{et al.} 2011), both relying on measurements of Shapiro delay,
which is not easily mimicked by other processes.

Also X-ray eclipses are now claimed to be an useful tool for determining masses of compact stars (Rawls \emph{et al.} 2011).
In previous studies the X-ray eclipse duration had been approximated analytically by assuming that the companion star is
spherical with an effective Roche lobe radius.  Now, to include all these various mass limits which are calculated with a fair degree of accuracy, it is almost impossible to fit in any single neutron star model. The problem with the gravitationally bound nuclear matter is that as we go to the lower mass values, the gravitational bound becomes weaker and the compact star tends to become larger in size. Whereas the exotic strange matter stars are bound by strong interaction as well as with gravity, and hence for a small mass, the radius is also small. Emphasis must be placed on the fact that some of these masses are not large
and only the strange star equation of state (EOS) can uniquely fit the masses of small and large compact objects. 

The idea of strange stars are well studied by now and in the model of Dey \emph{et al.} (1998) they were applied to 
4U 1820$-$30 and Her-X-1 claiming a more compact nature of stellar objects compared to neutron stars - and also strange
stars based on the MIT Bag model. The analysis of spectroscopic data on multiple thermonuclear bursts from 4U 1820$-$30 in
the globular cluster NGC 6624 yields well constrained values for the apparent emitting area and the Eddington flux, both of 
which depend in a distinct way on the mass and radius of the compact star. 

In the original model (Dey \emph{et al.} 1998) the masses of the star 4U 1820$-$30 and Her X$-$1 were estimated theoretically 
including different and appropriate scales for quark confinement and asymptotic freedom. They are now fitted to more accurate
observations in this present paper. We find that the mass and radius of 4U 1820$-$30 and 4U 1608$-$52  which is found from
new spectroscopic observations is fitted well with the new equation of state (EOS).  We also have fitted the large masses of 
PSR J1614$-$2230  and PSR J1903$-$0327 found from recent Shapiro delay observations and predict their radii. We have predicted the radii of the 
six eclipsing binary pulsars including Her X-1 using their masses fitted with current accurate determinations
based on Roche lobe geometry (Rawls \emph{et al.} 2011).

\section{The model and EOS}

The strange matter equation of state that we have chosen for our present work is based on the fact that there are interacting quarks and the quark masses follow an asymptotic behaviour. It is an extension of the strange matter EOS developed by Dey \emph{et al.} (1998), where there are interacting quarks, interaction taking place via the Richardson potential in a mean field Hartree-Fock prescription. The parameter set that we employ used the asymptotic  parameter $\Lambda$ = 100 MeV and confinement parameter $\Lambda^\prime$= 360 MeV along with an $\alpha_0$ = 0.43. To obtain larger masses for the strange stars we must have onset of chiral symmetry earlier with the free parameter N=2 since the lighter quarks can carry  more kinetic energy with the masses given as below 

\begin{equation} M_i = m_i + M_q {\rm sech}\left(\frac{n_B}{Nn_0}\right) , ~~~ i=u,d,s 
\label{eq:quark-mass}
\end{equation}

The parameter N dictates how fast the quark masses fall off with increasing density as is given in Eqn. (\ref{eq:quark-mass}). The confinement in the medium is softened and the corresponding inverse Debye screening length to the lowest order is given by (Baluni, 1978):

\begin{equation}
\rm (D^{-1})^2 \equiv \f{ 2 \al _0}{\pi} \sum_{\it i=u,d,s,}k^{\it f}_{\it i}
\sqrt{(k^{\it f}_{\it i})^2 + m_{\it i}^2} \label{eq:gm3}
\end{equation}
where ${\rm  k}^f_i$, the  Fermi momentum of  the {\it i}-th  quark is
obtained from the corresponding number density:
\begin{equation}
{{\rm k}^f_i} = (\rho _i \pi^2)^{1/3}
\end{equation}
and $\alpha_0$ is the perturbative quark gluon coupling. 

\begin{figure}
\centerline{\includegraphics[width=6cm,angle=-90]{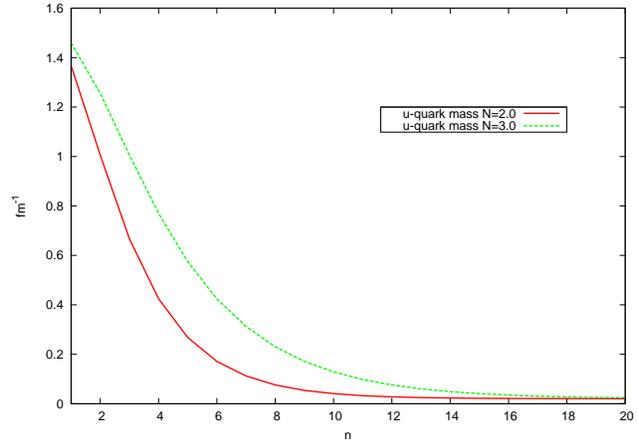}}
\caption{u-quark mass  for densities $n=n_B/n_0$. where $n_B$ is the baryonic density and $n_0$ is the normal nuclear matter density. Notable is faster fall off u- quark mass with increasing $n$ for N = 2 compared to N = 3 (Dey {\it et al.}, 1998)}
\label{fig:umass}
\end{figure}

Assuming chiral symmetry is restored at high energy, the quark mass, $M_i$, of i-th flavour, is taken to be density 
dependent so that it decreases with increasing density. $M_q$ is the quark mass when chiral symmetry is completely 
broken. $m_i$ are the current quark masses,  $m_u = 4~MeV$,  $m_d = 7~MeV$ and  $m_s = 150~MeV$ and the chiral symmetry restoration is shown in figure(\ref{fig:umass}) as a function of the baryonic density $$n_B = \frac{n_u + n_d + n_s}{3},$$ $n_0 = 0.17~fm^{-3}$. In the figure, we show two values of N, one corresponding to Dey \emph{et al.} (1998) and the other, N=2.0 taken in this paper. The argument for taking a lower value of N is that the masses  are decreasing faster as one goes inside the pulsar from the surface. That means there is a faster decrease in the scalar potential leading to sharper change in kinetic energy. The vector potential has not significantly changed from Bagchi \emph{et al.} (2006), but both the asymptotic part and the Debye screened part has relatively altered with sharper change in kinetic energy.

The detection of massive ($\sim 2 M_{\odot}$) pulsars has emphasized the need to study compact
astrophysical objects since they  place very strong constraints on the equation of state of
matter at extreme nuclear densities (e.g., Lattimer \& Prakash 2004).  Stars having hyperons or quark stars
having boson condensates, with softer EOS can barely reach such limits. In this paper we try to 
explain these heavy pulsars as strange stars (SS) with density dependent quark mass. We will show that
strange star models fit the the measured mass of pulsars ranging from $\sim 1 M_{\odot}$ to $\sim 2 M_{\odot}$,
because in these models the quark masses decrease fast with density as is shown in Figure~(\ref{fig:umass}).

\begin{table}
\caption{The parameters used for the EoS}
\vskip .3cm
\centering
\begin{tabular}{|c|c|c|c|c|c|c|c|}
\hline \hline
$\Lambda^\prime$ & $M_q$ & N & $\alpha_0$ & $(E/A)_{min}$ & $M_{max}$ & R \\
(MeV) & (MeV) & & & (MeV) & $M_\odot$ & (km)\\
\hline
360 & 400 & 2.0 & 0.43 & 728.12 & 2.0545 & 9.5 \\
\hline
\end{tabular}
\end{table}

The work of Li \emph{et al.} (1995) on Her X$-$1 and that of Bombaci (1997) led us to our realistic 
strange star model where the quark mass was taken to be density dependent and the interquark force had asymptotic freedom and 
confinement (with Debye screening) built into it through a Richardson type potential. The spirit is that of large colour model 
of t'Hooft where quark degrees of freedom are effectively approximated.

\section {The star Her X$-$1 }

Is Her~X-1 a strange star ? In 1995 Li, Dai and Wang first raised this question. They estimated its mass 
to be $0.98\pm0.12~ M_\odot$. In a more recent paper in 2008, Abubekerov \emph{et al.} (2008) reported the mass to be $0.85 \pm 0.15 ~ M_{\odot}$. In another recent article by Rawls \emph{et al.} (2011), they claim to have made  more refined mass determinations of six eclipsing stars, the Her X$-$1 mass is not very well determined with a large range $1.07 \pm 0.36~ M_{\odot}$. 

In our prediction for the radius, we have chosen the mass estimate of Abubekerov \emph{et al.} (2008) where they have deployed a physically justified technique of calculating the local absorption and taking into account external X-ray heating, and analyzing the radial-velocity curve  they
obtained the mass for the X-ray pulsar to be $0.85 \pm 0.15 ~ M_{\odot}$. The radius prediction for   Her X$-$1 from our EoS is around 8.1$ \pm 0.41$ km. For the numerical X-ray eclipsing calculations we expect more precise determinations will be made soon.

\begin{figure}
\centerline{\includegraphics[width=6cm,angle=-90]{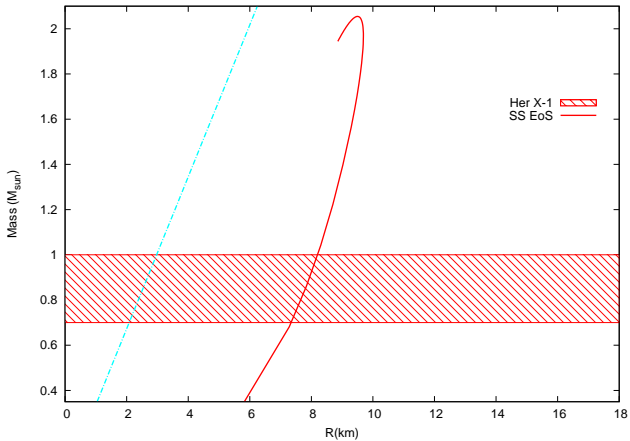}}
\caption{The refined mass estimate of Her X-1 infers a radius estimate of 8.1 $\pm$ 0.41 km from the SS model}
\label{fig:herX}
\end{figure}

\section{Mass measurement of 4U~1820$-$30, 4U~1608$-$52 and EXO~17454$-$248}

Simultaneously measuring the mass and radius of a distant compact star requires measurement of combination of several spectroscopic phenomena observed from the stellar surface. If the redshifted absorption line is absent from the spectroscopic feature, then it is extremely essential to know the distance of the source/object. In a recent work, G\"uver, Wroblewski, Camerata and \"Ozel (2010a) used time resolved X-ray spectroscopy of the thermonuclear burst of 4U 1820-30, and side by side they measured the distance of the Globular Cluster  NGC 6624, which hosts the same. They report well-constrained values of the apparent emitting area and the Eddington flux, both of which depend in a distinct way on the mass and radius of the compact star. They find the mass to be $M(4U 1820$-$30) = 1.58\pm 0.06 \; M_\odot$ and the radius to be $R = 9.1 \pm 0.4$ km. From our equation of state, we have a radius estimate of about 9.65 km assuming $M(4U 1820$-$30) = 1.58 M_\odot$ which is very close to their estimated value. Said so, they also mention that the relatively large uncertainty in the source distance estimate was taken care of using the Bayesian analysis of the probability density of a few set of values for the mass and radius pair for different distance values and optimising them for those set of values which fits well with the spectroscopic data, using a maximum likelihood method. 

In another work, G\"uver \emph{et al.} (2010b) measured the mass and radius of the lower mass X-ray binary (LMXB) 4U 1608$-$52. For this they used red clump giants as the standard candles to measure the distance of the LMXB and found the best fit distance estimate to be 5.8 kpc. Then they combined this distance estimate together with the time resolved spectroscopy of the Type I X-ray bursts to measure the mass and radius of the compact star in the X-ray binary,  by carrying out the similar likelihood analysis of the probability density functions of the mass and the radius, and found the optimised values to be  $M(4U 1608-52)=1.74\pm 0.14 \; M_{\odot}$, $R = 9.3 \pm 1.0$ km.  Again, comparing this with our estimated value of the radius for 1.74 $M_\odot$ star to be at 9.8 km, we observe that it falls well within the 1$\sigma$ contour of their observed value.

\"{O}zel \emph{et al.} (2009) used time resolved spectroscopic data from EXO~1745$-$248 during thermonuclear bursts, to measure the Eddington flux and apparent surface area of the compact star. EXO~1745$-$248 is located in the metal rich globular cluster Terzan 5, in the galaxy. A refined measurement of the distance of Terzan 5  using HST/NICMOS(Near Infra-red Camera and Multi Object Spectrometer) data revealed that this globular cluster is situated at a distance of 6.3 kpc, with an uncertainty of 10\%. The EXO~1745$-$248 was observed with the RXTE for 148 ks, where there were evidence of the photospheric radius expansion. Combining the distance estimate to the Terzan 5 and the photospheric radius expansion, they found a refined mass and radius estimate of 1.7 $M_\odot$ and 9 km respectively.

\begin{figure}
\centerline{\includegraphics[width=6cm,angle=-90]{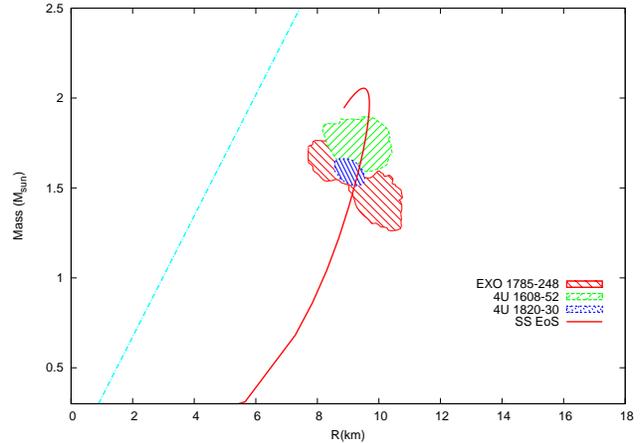}}
\caption{M-R plot with 4U 1820$-$30, 4U 1608$-$52 and EXO 1785-248 shows that within 1 $\sigma$ error, all of these compact objects agree with our SS model}
\label{fig:psr}
\end{figure}

\section{Mass measurement with Shapiro delay}

Shapiro delay is a general relativistic increase in light travel time through the curved space time near a massive body. For highly inclined (nearly edge on) binary millisecond radio (msr) pulsar system, this effect allows us to infer the masses of both the neutron star and its binary companion to high precision.
 
Demorest \emph{et al.} (2010) used the measured arrival times of the pulses to determine key physical parameters
of the compact star and its binary system by fitting them to a
comprehensive timing model that accounts for every rotation of the
star over the time spanned by the fit. The model predicts at
what times pulses should arrive at Earth, taking into account pulsar
rotation and spin-down, sky position, binary orbital parameters, time-variable interstellar dispersion
and general-relativistic effects such as the Shapiro delay.
They compared the observed arrival times with the model predictions,
and obtained best-fit parameters by $\chi^2$ minimization.

Two such determinations have already been made, the most accurate one to date being PSR~J1903$+$0327 with the mass 1.667 $\pm$ 0.021 $M_\odot$(Freire \emph{et al.}, 2011) and a more heavy msr PSR~J1614$+$2230 with mass 1.97 $\pm$ 0.04 $M_\odot$(Demorest \emph{et al.}, 2010). The radius of these stars still remain ambiguous and are dependent on the choice of the model/equation of state to describe them. 

The origin of the heavy PSR~J1614$+$2230 has been of considerable interest. Tauris, Langer and Kramer (2011) have looked at this star whose mass also was determined very accurately using Shapiro Delay method and concluded that this star must have originated with mass of either ~ 1.95 $M_\odot$ or 1.7 $\pm 0.15 M_\odot$ which according to them significantly exceeds birth masses of previously discovered pulsar systems and they suggested  that it was born massive from a progenitor star more massive than $20 M_\odot$.

To account for the high mass of PSR~J1614$+$2230, Demorest \emph{et al.} (2010) points out that such high mass can be supported by quark matter equation of state only if the quarks are interacting. In our refined strange matter and strange star model, where we have interacting quarks, we can see that if the compact star in PSR~J1614$+$2230 is an exotic strange matter star, then it should have a predicted radius of $\sim$ 9.5 km.

\begin{figure}
\centerline{\includegraphics[width=6cm,angle=-90]{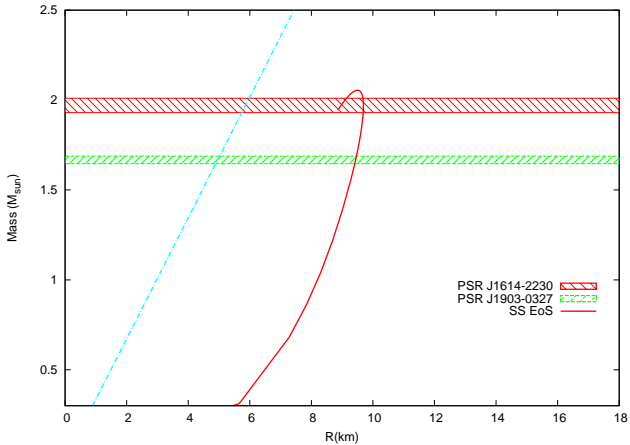}}
\caption{The SS model used in Figure (\ref{fig:psr}), also fits  PSR~J1903$+$0327, the highest  measured mass star, and PSR~J1614$+$2230, the most precise measurement mass of a millisecond pulsar.}
\label{fig:shapiro}
\end{figure}

There has also been excitement about the nature and evolution of the unique binary pulsar PSR~J1903$+$0327,   which has a mass $1.667 \pm 0.021~M_\odot$ (0.99 \% confidence limit). Freire \emph{et al.} (2011) have suggested that it is a black widow which accreted up its nearest partner fully or partially so that it disappeared from a tertiary system after it was formed from a supernova. They have further suggested that its rather large eccentricity is relatively constant.

\section{Comparing with 6 eclipsing binary pulsars}
Rawls \emph{et al.} (2011) have refined mass determination for six eclipsing X-ray pulsar binaries which fit into our present investigation of strange star candidates. Of these the Her X-1 has a  mass with big errors : $1.07 \pm 0.36 M_\odot$ but  was the first star suggested to be strange on the basis of its radius estimate. 

Rawls \emph{et al.} (2011) now use a numerical code based on Roche geometry with various optimizers to analyse the published data for these systems, which we supplement with new spectroscopic and photometric data for 4U 1538-52. This allows them to model the eclipse duration more accurately and thus calculate an improved value for the neutron star mass. The derived neutron star mass also depends on the assumed Roche lobe filling factor $\beta$ of the companion star, where $\beta$ = 1 indicates a completely filled Roche lobe. In previous work a range of $\beta$ between 0.9 and 1.0 was usually adopted. Now optical ellipsoidal light-curve data is used to constrain $\beta$. Rawls \emph{et al.} (2011) find neutron star masses of $1.77 \pm 0.08 M_\odot$ for Vela X$-$1, $0.87 \pm 0.07 M_\odot $ for 4U 1538$-$52 (eccentric orbit), $1.00 \pm 0.10 M_\odot $ for 4U 1538$-$52 (circular orbit), $1.04 \pm 0.09 M_\odot $ for SMC X$-$1, $1.29 \pm 0.05 M_\odot$ for LMC X-4, $1.49 \pm 0.08  M_\odot$ for Cen X-3, and $1.07 \pm 0.36 M_\odot $ for Her X-1.

The earlier methods relied on analytic approximations for the
Roche lobe geometry, whereas
 ELC code of Jerry Orosz does not rely on such approximations but rather
solves the Roche lobe geometry numerically. In figures 2-4  of Rawls \emph{et al.} (2011), they
explore the parameter space and show that the analytic approximations
can be off by as much as 10 to 20\% in some cases. Figures 5 and 6 of Rawls \emph{et al.} (2011)
again show how much the shape of the Roche lobe filling donor star can
differ based on whether you use analytic approximations (red lines) or
their numerical code (solid black lines).

\begin{figure}
\centerline{\includegraphics[width=6cm,angle=-90]{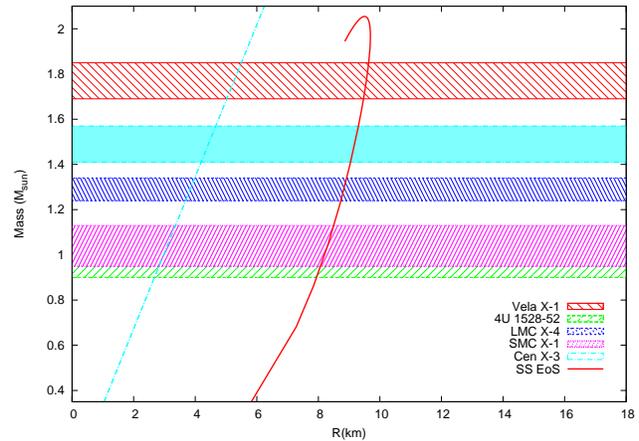}}
\caption{The measured mass of the 5 X-ray binary pulsars with OB Supergiant Companion using Roche geometry, agree with our SS model - predicting their radii between 8-10 km}
\label{fig:5puls}
\end{figure}

The important things from their paper are the results
presented in Table 4 and visually summarized in Fig. 10.  The solid
lines in Fig. 10 are their 'improved' mass estimates, and the dotted
lines are what you would get by older/analytic methods.  As we can
see, in most cases the error bars for their new/numerical method are
indeed smaller than those derived via analytic method.  But other than
4U~1528$-$52 the change in compact star masses (derived by the two methods)
are quite small.  For 4U~1528$-$52 it appears that their numerical method
gives two (almost significantly) different results whether one assumes
the orbit to be eccentric (then one gets a lower NS mass) or circular
(then one gets a bit higher NS mass). Current data cannot distinguish
between eccentric or circular orbits for this one.

\section{Ramifications from PSR J1614-2230, Vela X-1 for heavy compact stars with heavier partners.}

Mass determinations of Vela X-1 (Rawls \emph{et al.}, 2011) suggest that this compact star has a high observed mass. 
The companion star to Vela X-1 is a B0.5 Ib supergiant (HD 77581) with a mass of about 23 M$_\odot$ which implies that the
present mass of the neutron star is very close to its birth-mass as suggested by Tauris (2011). Even a hypothetical strong 
wind accretion at the Eddington limit would not have resulted in accretion of more than about $10^{-2}$ M$_\odot$ given the short 
lifetime of its massive companion. We therefore conclude that not only was the compact star in PSR~J1614$-$2230 born massive
$1.97 ~\pm~0.04~M_\odot$, but also the compact star Vela X-1 was born with a mass  $1.77~\pm0.08~M_\odot$. In these cases 
it is possible that the pulsars did not have time to go over to the paired phase. Even if they did we checked that the masses 
would not change very much but the radius will decrease - in other words the stars will be more compact. In what follows we 
are trying to show that all the 11 compact stars above, some with high masses and some lower, can be explained with a single strange 
star EoS.

\section{The compact star XTE J1739$-$285 and SAX~J1808.4$-$3658}

Zhang, Yin, Zhao, Wei and Li (2007) considered the star XTE~J1739$-$285 which is very interesting since its spinning 
period may be as high as 1122 Hz. The uncertainty in the determination of the spin remains a problem since it is observed 
in only one burst but not in the whole set of bursts (Kaaret \emph{et al.}, 2007). In Zhang \emph{et al.} (2007), 
the authors suggested that if the higher twin QPOs are detected in XTE~J1739$-$285, a strong constraint on mass and radius can be 
placed using the upper kHz quasi-periodic oscillation frequency and the star frequency one gets  $M~=~1.51 M_\odot$ and 
radius $<$ 10.9 km and this excludes most EoS of normal neutron matter. They observed that this is fitted well with EoS 
SS1 given in Dey \emph{et al.} (1998). The present SS EoS used here is a refined and modified form of the Strange Star EoS (SS1) of Dey \emph{et al.} (1998).

The SAX~J1808.4$-$3658 has a mass of  $0.9 \pm 0.3 M_\odot$ from optical measurements (Elebert \emph{et al.}, 2009 ) but according to our QPO study (Gangopadhyay \emph{et al.}, 2012) for this star the central value of the mass is preferred. 

Further the precise orbital measurements have prompted Di Salvo \emph{et al.}  (2008) and Burderi \emph{et al.} (2009) to suggest that the orbit enhancement and the estimated accretion from the luminosity of the X-ray bursts over ten years suggest that  there is unconventional mass loss from the binary system. Electrons form a layer outside SS since they are held by electromagnetism, as opposed to quarks held by strong interaction. These electrons will interact with the u-quarks of the incoming accreting particles (normal matter). Hence, in terms of the conversion of the normal matter from the  brown dwarf at the surface of the strange star the reaction is simply: 

$$u + e^- = s + \nu$$

since the gain in  energy in our EoS at the surface of the strange star is 200 MeV whereas the mass excess of s over u is only 150 MeV. 
This strongly supports the conclusions of Li \emph{et al.} (1999) suggesting that SAX~J1808.4$-$3658 is a strange star.

Further confirmation in future burst detections is awaited hopefully with detection of higher twin QPOs. In a recent paper we have shown that for many stars with measured higher twin QPOs including SAX~J1808.4$-$3658 one can indeed predict 
the mass and radius for fast rotating stars using Kerr geometry results (Gangopadhyay \emph{et al.}, 2012). 

\begin{table*}
\centering
\caption{Revised radii prediction for 4U 1820-30, 4U1608-52, SAX J1808.4-3658 and EXO 1785-248, and new radii prediction for the other 8 stars.}
\vskip .3cm
\begin{tabular}{|l|c|c|c|c| }
\hline
STAR & $M$ & R & R(predicted)& References\\
& ($M_\odot$) & (km)& (km)&\\
\hline
PSR J1614-2230 & 1.97 $\pm$ 0.04 & - & 9.69 $\pm$ 0.2 & Demorest \emph{et al.} (2010) \\
PSR J1903+327 & 1.667 $\pm$ 0.021 & - & 9.438 $\pm$ 0.03 &  Freire \emph{et al.} (2011)\\
Vela X-1 & 1.77 $\pm$ 0.08 & - & 9.56 $\pm$ 0.08  & Rawls \emph{et al.} (2011) \\
4U 1538-52 & 0.87 $\pm$ 0.07 & - & 7.866 $\pm$ 0.21& Rawls \emph{et al.} (2011) \\
LMC X-4 & 1.04 $\pm$ 0.09 & - & 8.301 $\pm$ 0.2 & Rawls \emph{et al.} (2011) \\
SMC X-4 & 1.29 $\pm$ 0.05 & - & 8.831 $\pm$ 0.09&  Rawls \emph{et al.} (2011) \\
Cen X-3 & 1.49 $\pm$ 0.08 & - & 9.178 $\pm$ 0.13 & Rawls \emph{et al.} (2011) \\
Her X-1 & 0.85 $\pm$ 0.15 & - & 8.1 $\pm$ 0.41 & Abubekerov \emph{et al.} (2008) \\
4U 1820-30 & 1.58 $\pm$ 0.06 & 9.1 $\pm$ 0.4 & 9.316 $\pm$ 0.086 & G\"{u}ver \emph{et al.} (2010a)\\
4U1608-52 & 1.74 $\pm$ 0.14 & 9.3 $\pm$ 1.0 & 9.528 $\pm$ 0.15 & G\"{u}ver \emph{et al.} (2010b)\\
SAX J1808.4-3658 & 0.9 $\pm$ 0.3 & - & 7.951 $\pm$ 1.0 & Elebert \emph{et al.} (2009)\\
EXO 1785-248 & 1.3 $\pm$ 0.2 & 10-12 & 8.849 $\pm$ 0.4  & \"{O}zel \emph{et al.} (2009)\\
\hline
\end{tabular}
\end{table*}

\section{Discussion}

Four stars Her X-1, 4U 1820-30, SAX J1808.4 and more recently the fast spinning XTE J1739-285 were suggested to be 
strange stars. We add the stars whose masses are recently determined accurately: two from Shapiro Delay, six from 
Roche Lobe geometry for eclipsing stars which includes Her X-1 and three from spectroscopic studies including 4U 1820$-$30. 
Fitting these stars with a M-R curve of strange star effectively restricts the radii of these stars within sharp limits since 
the nature of the curve is restricted to smaller radii compared to neutron stars. As is well known the reason is that neutron 
stars have to be bound by gravitation so that to have a small radius it has to have a large mass. Strange stars on the other 
hand are bound by strong interaction as well as gravitation so that small mass stars have necessarily small radii and for the 
large mass stars also the radius is not too large - as can indeed be seen from our figures 2, 3 and 4. 
The discovery of a 2$M_\odot$ binary millisecond pulsar raises the interesting
possibility that PSR J1614$-$2230 was born from the collapse of a very massive 29 $M_\odot$ as discussed by Tauris \emph{et al.} (2011). This would support
the idea of it being a strange star since this might lead to the strange matter phase. There is problem regarding the origin of 
the other Shapiro-delay star PSR J1903+0327 and Freire \emph{et al.} (2011) comment that stellar evolution studies may provide 
new insights on how millisecond pulsars form.
In a recent paper Logoteta, Providencia, Vidana and Bombaci (2012) find that it is difficult to obtain masses larger than 
1.62 $M_\odot$ even with hyperon or hybrid stars.

\section*{Acknowledgements}

The work of JD and MD was supported in part from a grant from the Department of Science and Technology, Govt. of India, 
New Delhi, no. SR/S2/HEP-026/2009. JD acknowledges HRI, Allahabad, India for short term visits.  JD and MD also acknowledge a long and useful discussion with Professeors Dmitri Psaltis and Feyal \"Ozel during a visit to the CFA in Harvard. SR acknowledges the 
India-South Africa bilateral research grant and the NRF incentive funding for research support.





\end{document}